\documentclass{ws-mpla}

\begin{document}

\markboth{Yu-Xiao Liu etc.}
{Energy-momentum for Randall-Sundrum
models}

\catchline{}{}{}{}{}

\title{Energy-momentum for Randall-Sundrum models}

\author{Yu-Xiao Liu\footnote{Corresponding author.}}
\address{Institute of Theoretical Physics, Lanzhou University,\\
   Lanzhou 730000, P. R. China\\
   liuyx@lzu.edu.cn}
\author{Li-Jie Zhang}
\address{Department of Mathematics and Physics,
   Dalian Jiaotong University,\\
   Dalian 116028, P. R. China}
\author{Yong-Qiang Wang}
\address{Zhejiang Institute of Modern Physics,
   Department of Physics, Zhejiang University,\\
   Hangzhou 310027, P. R. China}
\author{Yi-Shi Duan}
\address{Institute of Theoretical Physics, Lanzhou University,\\
   Lanzhou 730000, P. R. China}

\maketitle


\begin{abstract}
We investigate the conservation law of energy-momentum for
Randall-Sundrum models by the general displacement transform. The
energy-momentum current has a superpotential and are therefore
identically conserved. It is shown that for Randall-Sundrum
solution, the momentum vanishes and most of the bulk energy is
localized near the Planck brane. The energy density is $
\varepsilon = \varepsilon_0 e^{-3k \mid y \mid}$.

\keywords{Energy-momentum; Randall-Sundrum models.}
\end{abstract}

\ccode{PACS Nos.: 04.20.Cv; 04.20.Fy; 04.50.+h.}


\section{Introduction}

The conservation law of energy-momentum (or the definition of
energy-momentum density) for the gravitational field is one of the
most fundamental and controversial problems in general relativity.
As a true field, it would be natural to expect that gravity should
has its own local energy-momentum density. However, it is usually
asserted that such a density can not be locally defined because of
the equivalence principle.\cite{Misner1973} As a consequence, many
attempts to identify an energy-momentum density for the
gravitational field lead to complexes that are not true tensors.
The first of such attempt was made by Einstein who proposed an
expression for the energy-momentum density of the gravitational
field. Bauer and Shr\"{o}dinger \cite{Bauer1918,Shrodinger1956}
pointed out that Einstein's expression of the energy-momentum
density has a serious shortcoming, i.e., it holds true only in the
quasi-Galilean coordinate system and that the transformation of
the description of flat space from rectangular coordinates to
spherical coordinates results in a nonzero energy density which
yields an infinite total energy. Indeed it was nothing but the
canonical expression obtained from N\"{o}ether's theorem
\cite{Trautman1962} and this quantity is a pseudotensor, an object
that depends on the coordinate system. Among the class of
conserved quantities, the Landau-Lifshitz
\cite{Landau-Lifshitz1951} quantity is the only symmetric one. The
advantage of symmetry is that it is easy in this case to construct
further conservation laws in invariant integral form, which in
turn are related to the angular momentum. But this did not
overcome the difficulty of Einstein's complex and was also only
admissible in quasi-Galilean coordinates. So Landau's expression
also has the same defects as Einstein's.

One of the present authors (Duan) had proposed one form of the
conservation law of energy-momentum in Riemann space
time,\cite{Duan1963} in which the energy-momentum has a Lorentz
index and a Riemann index, and hence is a covariant vector in
Riemann space time. This conservation law is general covariant and
it overcomes the flaw of the Einstein and Landau
\cite{Landau-Lifshitz1951,Einstein1916} forms of the
energy-momentum conservation law.
It can explain the nonzero energy flux density and energy density
for Bondi's plane wave, and get the correct gravitational
radiation formula,\cite{Duan1983} where one could not get the
nonzero energy density for Bondi's plane wave by M{\o}ller's
formula.\cite{Moller1958,Moller1961,Moller1963} This conservation
law had been generalized to the general space
time.\cite{Duan1987,Duan1988}

On the other hand, Theories with extra dimensions have recently
attracted enormous attention. The possible existence of such
dimensions got strong motivation from theories that try to
incorporate gravity and  gauge interactions in a unique scheme, in
a reliable manner. The idea dates back to the 1920's,  to the
works of Kaluza and Klein \cite{Kaluza,Klein} who tried to unify
electromagnetism with Einstein gravity by assuming that the photon
originates from the fifth component of the metric. In the course
of the last several years, there has been active interest in the
brane world scenarios
\cite{Arkani-Hamed,AADD1998,CremadesNPB2002,KokorelisNPB2004,RSI,RSII,GarrigaPRL2000,GiddingsJHEP2000,LykkenJHEP2000,ShiromizuPRD2000}
and fermionic zero modes in Large
dimensions.\cite{FrereJHEP2003,WangMPLA2005,LiuCTP2007,LiuJHEP2007,LiuNPB2007}
The pioneering work was done by Randall and
Sundrum.\cite{RSI,RSII} In their works, they present the so called
Randall-Sundrum (RS) models
\cite{RSI,RSII,GogberashviliIJMPD2002,ArkaniPRL2000} for warped
backgrounds, with compact or even infinite extra dimensions. The
RSI scenario provides a way to solve the hierarchy problem, and
the RSII scenario gives Newton's law of gravity on the brane of
positive tension embedded in an infinite extra dimension.

As in the (3+1)-dimensional case, we should have some conservation
laws in order to understand high dimensional gravity well. In this
paper, we would like to study the simple but typical RS models.
The purpose of this paper is to present the relationship between
conservation theorems and invariance properties of physical
systems in the models. It is the extension of previous works and
includes the original general relativity formula of Einstein. In
the present paper we pay more attention to the energy-momentum
conservation law based on the Lagrangian which includes the
contribution of the branes in RS models, and prove that there
exists a superpotential with respect to the Lagrangian.

The paper is arranged as follows. In section
\ref{RSConservationII}, we give a general description of the
scheme for establishing general covariant conservation laws in
general relativity. In section \ref{RSConservationIII}, we first
give a simple review of the general displacement transform and the
RS models, then use the general displacement transform and the
scheme to obtain a general covariant conservation law of
energy-momentum for RS models. At the last of this section, we
calculate the energy density and the total energy and momentum of
the bulk for Randall-Sundrum solution by the superpotential.
Section \ref{RSConservationIV} is devoted to some remarks and
discussions.

\section{Conservation laws in general relativity}\label{RSConservationII}
The conservation law is one of the important problems in
gravitational theory. It is due to the invariance of the action
corresponding to some transforms. In order to study the covariant
energy-momentum law of more complicated systems, it is necessary
to discuss conservation laws by N\"{o}ether theorem in the general
case.\cite{Duan1963,Duan1987,Duan1988,Feng1999,Cho1995} Suppose
that the space-time manifold $\cal{M}$ is of dimension $n=1+d$ and
the Lagrangian density is in the first order formalism, i.e.
\begin{equation} \label{action}
I=\int_{\cal M} d^{n}x {\cal{L}}(\phi^{A},
{\partial}_{\mu}\phi^{A}),
\end{equation}
where $\phi^{A}$ denotes the general fields. If the action is
invariant under the infinitesimal transformations
\begin{eqnarray}
x^{\prime\mu} &=& x^{\mu}+\delta x^{\mu}, \label{transformation1}\\
\phi^{\prime A}(x^{\prime}) &=&
\phi^{A}(x)+\delta\phi^{A}(x),\label{transformation2}
\end{eqnarray}
and $\delta\phi^{A}$ vanishes on the boundary of $\cal M$,
$\partial \cal M$, then following relation holds
\cite{Duan1963,Duan1987,Duan1988,Feng1995}
\begin{equation}
 \partial_{\mu}\left({\cal{L}}{\delta} x^{\mu}
       +\frac{{\partial}{\cal{L}}}
             {{\partial}{\partial}_{\mu}\phi^{A}}
        \delta_{0}\phi^{A} \right)
 +[{\cal{L}}]_{\phi^{A}}\delta_{0}\phi^{A}=0, \label{NoetherTheorem}
\end{equation}
where
\begin{equation}
[{\cal{L}}]_{\phi^{A}}=\frac{{\partial}{\cal{L}}}{{\partial}\phi^{A}}-{\partial}_{\mu}
\frac{{\partial}{\cal{L}}}{{\partial}{\partial}_{\mu}\phi^{A}},
\end{equation}
and $\delta_{0}\phi^{A}$ is the Lie derivative of $\phi^{A}$
\begin{equation}
\delta_{0}\phi^{A}=\phi^{\prime A}(x)-\phi^{A}(x)=
\delta\phi^{A}(x)-{\partial}_{\mu} \phi^{A}\delta x^{\mu}.
\end{equation}

If ${\cal{L}}$ is the total Lagrangian density of the system, the
field equation of $\phi^{A}$ is just $[{\cal{L}}]_{\phi^{A}}=0$.
Hence from Eq. (\ref{NoetherTheorem}), we can obtain the
conservation equation corresponding to transformations
(\ref{transformation1}) and (\ref{transformation2})
\begin{equation} \label{ConservationEq}
 {\partial}_{\mu}\left({{\cal{L}}}{\delta} x^{\mu}
 +\frac{{\partial}{\cal{L}}}{{\partial}{\partial}_{\mu}\phi^{A}}
\delta_{0} \phi^{A}\right)=0.
\end{equation}
It is important to recognize that if ${\cal{L}}$ is not the total
Lagrangian, e.g. the gravitational part ${\cal{L}}_{g}$, then so
long as the action of ${\cal{L}}_{g}$ remains invariant under
transformations (\ref{transformation1}) and
(\ref{transformation2}), Eq. (\ref{NoetherTheorem}) is still valid
yet Eq. (\ref{ConservationEq}) is no longer admissible because of
$[{\cal{L}}_{g}]_{\phi^{A}}\not=0$.

In a gravitational theory with the vierbein as elementary fields,
we can separate $\phi^{A}$ as $\phi^{A}=(e^{\mu}_{a}, \psi^{B})$,
where $\psi^{B}$ is an arbitrary tensor under general coordinate
transformations. Suppose that ${\cal{L}}_g$ does not contain
$\psi^{B}$, then Eq. (\ref{NoetherTheorem}) reads
\begin{equation} \label{NoetherTheoremOfLg1}
{\partial}_{\mu} \left( {\cal L}_{g}\delta x^{\mu}
    +\frac{{\partial}{\cal{L}}_{g}}{{\partial}
     {\partial}_{\mu}e^{\nu}_{a}} \delta_{0}e^{\nu}_{a} \right)
    +[{\cal L}_{g}]_{e^{\nu}_{a}}\delta_{0}e^{\nu}_{a}=0.
\end{equation}
Under transformations (\ref{transformation1}) and
(\ref{transformation2}), the Lie variations are
\begin{equation} \label{LieVariationOfVierbein}
\delta_{0}e^{\mu}_{a} = e^{\nu}_{a} \delta x^{\mu}_{,\nu}
-e^{\mu}_{a,\nu}\delta x^{\nu},
\end{equation}
where ``," denotes partial derivative. Substituting Eq.
(\ref{LieVariationOfVierbein}) into Eq.
(\ref{NoetherTheoremOfLg1}) gives
\begin{eqnarray}
{\partial}_{\mu} \left[ \left( {\cal L}_{g}\delta^{\mu}_{\sigma}
-\frac{{\partial}{\cal{L}}_{g}}{{\partial}e^{\nu}_{a,\mu}}
e^{\nu}_{a,\sigma} \right) \delta x^{\sigma} +
\frac{{\partial}{\cal{L}}_{g}}{{\partial}e^{\nu}_{a,\mu}}
e^{\sigma}_{a} \delta x^{\nu}_{,\sigma} \right]
 + [{\cal{L}}_{g}]_{e^{\mu}_{a}}(e^{\nu}_{a} \delta x^{\mu}_{,\nu} -
e^{\mu}_{a,\nu} \delta x^{\nu})=0.~~\label{NoetherTheoremOfLg2}
\end{eqnarray}
Comparing the coefficients of $\delta x^{\mu}, \delta
x^{\mu}_{,\nu} $ and $\delta x^{\mu}_{,\nu\lambda}$, we can obtain
an identity
\begin{equation}
[{\cal{L}}_{g}]_{e^{\nu}_{a}}e^{\nu}_{a,\mu} +
{\partial}_{\nu}([{\cal{L}}_{g}]_{e^{\mu}_{a}}e^{\nu}_{a})=0.
\end{equation}
Then Eq. (\ref{NoetherTheoremOfLg2}) can be written as
\begin{eqnarray}
{\partial}_{\mu} \left[ \left( {\cal{L}}_{g}\delta^{\mu}_{\sigma}
-\frac{{\partial}{\cal{L}}_{g}}{{\partial}e^{\nu}_{a,\mu}}
e^{\nu}_{a,\sigma} + [{\cal{L}}_{g}]_{e^{\sigma}_{a}} e^{\mu}_{a}
\right) \delta x^{\sigma} 
 + \frac{{\partial}{\cal{L}}_{g}}{{\partial}e^{\nu}_{a,\mu}}
e^{\sigma}_{a} \delta x^{\nu}_{,\sigma} \right]=0.
\label{ConservationLaw1}
\end{eqnarray}
This is the general conservation law in the vierbein formalism of
general relativity. By definition, we introduce
\begin{eqnarray}
\tilde{I}^{\mu}_{\sigma}~&=&{\cal{L}}_{g}\delta^{\mu}_{\sigma}
-\frac{{\partial}{\cal{L}}_{g}}{{\partial}e^{\nu}_{a,\mu}}
e^{\nu}_{a,\sigma} + [{\cal{L}}_{g}]_{e^{\sigma}_{a}}
e^{\mu}_{a}, \label{Imusigma} \\
\tilde{V}^{\mu\sigma}_{\nu}&=&\frac{{\partial}{\cal{L}}_{g}}{{\partial}e^{\nu}_{a,\mu}}
e^{\sigma}_{a}. \label{Vmusigmanu}
\end{eqnarray}
Then Eq. (\ref{ConservationLaw1}) gives
\begin{equation}\label{ConservationLaw2}
{\partial}_{\mu}(\tilde{I}^{\mu}_{\sigma}\delta x^{\sigma}+
\tilde{V}^{\mu\sigma}_{\nu} \delta x^{\nu}_{,\sigma})=0.
\end{equation}
Eq. (\ref{ConservationLaw2}) is tenable under arbitrary
infinitesimal transformations, so we can compare the coefficients
of $\delta x^{\sigma}, \delta x^{\sigma}_{,\mu}$ and $\delta
x^{\sigma}_{,\mu\lambda}$ and obtain
\begin{eqnarray}
{\partial}_{\mu}\tilde{I}^{\mu}_{\sigma}&=&0,\label{Result1}\\
\tilde{I}^{\mu}_{\sigma}~&=&-{\partial}_{\nu}\tilde{V}^{\nu\mu}_{\sigma},\label{Result2}\\
\tilde{V}^{\mu\sigma}_{\nu}&=&-\tilde{V}^{\sigma\mu}_{\nu}.\label{Result3}
\end{eqnarray}
Eqs. (\ref{Result1})-(\ref{Result3}) are fundamental to the
establishing of conservation law of energy-momentum.

\section{Conservation law of energy-momentum for
RS models}\label{RSConservationIII}

\subsection{General displacement transformations}
In 3+1 dimension, the conservation of energy-momentum in special
relativity is a consequence of the invariant property of the
action under the infinitesimal translation of the Lorentz
coordinates
\begin{equation}
x^{\prime \bar{a}}=x^{\bar{a}}+b^{\bar{a}}, \;\;\;\;
b^{\bar{a}}=const. \;\;\;\;(\bar{a}=0,1,2,3)
\end{equation}
The corresponding transformation of arbitrary coordinates in flat
spacetime is
\begin{equation}
x^{\prime \bar{\mu}}=x^{\bar{\mu}}+\delta x^{\bar{\mu}},
\;\;\;\;(\bar{\mu}=0,1,2,3)
\end{equation}
where
\begin{equation}
\delta
x^{\bar{\mu}}=\frac{{\partial}x^{\bar{\mu}}}{{\partial}x^{\bar{a}}}\delta
x^{\bar{a}}=\frac{{\partial}x^{\bar{\mu}}}{{\partial}x^{\bar{a}}}b^{\bar{a}}.
\end{equation}
If we extend this idea to general relativity, we have the
generalized translation transformation \cite{Duan1963}
\begin{equation} \label{translation}
x^{\prime \bar{\mu}}=x^{\bar{\mu}}+\delta x^{\bar{\mu}}, \;\;\;\;
\delta x^{\bar{\mu}}=e^{\bar{\mu}}_{\bar{a}}b^{\bar{a}}.
\end{equation}
Using the invariance of the action with respect to {\em general
displacement transformations} (\ref{translation}) and Einstein
equations, we get the following general covariant conservation law
of energy-momentum
\begin{equation}\label{4DConservationLaw}
\nabla_{\bar{\mu}}(T^{\bar{\mu}}_{a}+t^{\bar{\mu}}_{\bar{a}})=0.
\end{equation}
The total energy-momentum is
\begin{equation}
P_{\bar{a}}=\frac{1}{c}\int_{\Sigma}d\Sigma_{\bar{\mu}}
\sqrt{-g}\left(T^{\bar{\mu}}_{\bar{a}} +
t^{\bar{\mu}}_{\bar{a}}\right)
=\frac{1}{c}\int_{S}dS_{\bar{\mu}\bar{\nu}}
V^{\bar{\mu}\bar{\nu}}_{\bar{a}},
\end{equation}
where $V^{\bar{\mu}\bar{\nu}}_{\bar{a}}$ is the superpotential
\begin{equation}
V^{\bar{\mu}\bar{\nu}}_{\bar{a}}=\frac{c^{4}}{8\pi G}
\left[e^{\bar{\mu}}_{\bar{b}}e^{\bar{\nu}}_{\bar{c}}
\omega_{\bar{a}}\,^{\bar{b}\bar{c}}
+\left(e^{\bar{\mu}}_{\bar{a}}e^{\bar{\nu}}_{\bar{b}}
-e^{\bar{\mu}}_{\bar{b}}e^{\bar{\nu}}_{\bar{a}}\right)
\omega^{\bar{b}} \right] \sqrt{-g}.
\end{equation}
This definition of energy-momentum has the following main
properties:

1). It is a covariant definition with respect to general
coordinate transformations. But the energy-momentum tensor is not
covariant under local Lorentz transformations, this is reasonable
because of the equivalence principle.

2). For a closed system, the total energy-momentum does not depend
on the choice of Riemann coordinates and transformations in the
covariant way
\begin{equation}
P_{\bar{a}}^{\prime}=L_{\bar{a}}\,^{\bar{b}}P_{\bar{b}}
\end{equation}
under local Lorentz transformation $\Lambda^{\bar{a}}\,_{\bar{b}}$
which is constant $L^{\bar{a}}\, _{\bar{b}}$ at spatial infinity.

3). For a closed system with static mass center, the total
energy-momentum is $P_{\bar{a}}=(Mc,0,0,0)$, i.e., the total
energy $E=Mc^{2}$.

4). For a rather concentrated matter system, the gravitational
energy radiation is \cite{Duan1983}
\begin{equation}
-\frac{{\partial}E}{{\partial}t} =\frac{G}{45c^{5}}
(\stackrel{\cdots}{D_{\bar{i}\bar{j}}})^{2},
\end{equation}
where $D_{\bar{i}\bar{j}}$ is the mass quadrupole moment. This
quadrupole radiation formula is in good agreement with the
observational data from the synthetic analysis of the
gravitational radiation damping for the pulsar PSR 1913+16 in a
binary star system.\cite{Taylor1979}

5). For Bondi's plane wave, the energy current is \cite{Duan1983}
\begin{equation}
t^{\bar{\mu}}_{0}=(\frac{1}{4\pi}\beta^{\prime 2},
\frac{1}{4\pi}\beta^{\prime 2}, 0,0),
\end{equation}
where $\beta$ is a function of $(t-x^1)$. From the above
expression we can see that the energy density is determined by
$\beta^{\prime 2}$ which always has positive values. While using
M{\o}ller's expression, Kuchar and Langer calculated the energy
density for Bodi's plane wave, the result obtained is zero.

6). For the solution of gravitational solitons, we can obtain
finite energy while the Landau-Lifshitz definition leads to
infinite energy.\cite{Yan1987}

7). In Ashtekar's complex formalism of general relativity, the
energy-momentum and angular momentum constitute a 3-Poincare
algebra and the energy coincides with the ADM
energy.\cite{Feng1995}

In the next subsection, we first give a brief introduction of the
RS models. Then, with these foundations above, we use
(4+1)-dimensional transformations (\ref{translation}) to obtain
conservative energy-momentum for RS models.

\subsection{RS models}

Let us consider the following setup. A five dimensional spacetime
with an orbifolded fifth dimension of radius $r$ and coordinate
$y$ which takes values in the interval $[0,\pi r]$. Consider two
branes at the fixed (end) points $y=0,\pi r$; with tensions $\tau$
and $-\tau$ respectively. The brane at $y=0$ ($y=\pi r$) is
usually called the hidden (visible) or Planck (SM) brane. We will
also assign to  the bulk  a negative cosmological constant
$-\Lambda$. Here we shall assume that all parameters are of the
order the Planck scale.

The classical action describing  the above setup is given by
\begin{equation} \label{ActionOfRS}
S_{g}= S_{0} + S_{h} + S_{v},
\end{equation}
here
\begin{equation}
S_{0} = \int \! d^4x \, dy \sqrt{g} \left( \frac{1}{2k_\star^2} R
+\Lambda\right)
\end{equation}
gives the bulk contribution, whereas the visible and hidden brane
parts are given by
\begin{equation}
S_{v,h}= \pm~\tau\int d^4x \sqrt{-g_{v,h}}~,
\end{equation}
where $g_{v,h}$ stands for the induced metric at the visible and
hidden branes, respectively. And $2k_\star^2=8\pi
G_\star=M_\star^{-3}$. Five dimensional  Einstein equations for
the given action are
 \begin{eqnarray} \label{eers}
 G_{\mu\nu} =-\; k_\star^2\Lambda\,g_{\mu\nu} +
                k_\star^2\tau\, \sqrt{\frac{-g_{h}}{g}}
                \delta_{\mu}^{\bar{\mu}} \delta_{\nu}^{\bar{\nu}}
                g_{\bar{\mu}\bar{\nu}}\delta(y)
       - k_\star^2\tau\, \sqrt{\frac{-g_{v}}{g}}
       \delta_{\mu}^{\bar{\mu}} \delta_{\nu}^{\bar{\nu}}
       g_{\bar{\mu}\bar{\nu}}\delta(y-\pi r),
 \end{eqnarray}
 where the Einstein tensor $G_{\mu\nu} = R_{\mu\nu} - \frac{1}{2} g_{\mu\nu}
 R$ as usual, Greek indices without bar $\mu, \nu = 0, \cdots, 4$
 and the others with bar $\bar{\mu}, \bar{\nu} = 0, \cdots, 3$.
 The solution that gives a flat induced metric on the branes is
\begin{equation}\label{rsmetric}
ds^2 = g_{\mu\nu}dx^\mu dx^\nu = e^{-2k
|y|}\eta_{\bar{\mu}\bar{\nu}}dx^{\bar{\mu}} dx^{\bar{\nu}} -
dy^2~,
\end{equation}
in which $x^{\bar{\mu}}$ are coordinates for the familiar four
dimensions, $k$ is a scale of order the Planck scale
\begin{equation} \label{rsmu}
k^2 =\frac{k_\star^2\Lambda}{6}= \frac{\Lambda}{6 M_\star^3}~,
\;\;\; \Lambda = \frac{\tau^2}{6 M_\star^3}~.
\end{equation}
The effective Planck scale in the theory is given by
\begin{equation}\label{kMpM*}
 M_{P}^2 = \frac{M_\star^3}{k}\left(1- e^{-2k\pi r}\right).
\end{equation}
Notice that for large $r$, the exponential piece becomes
negligible, and above expression has the familiar form given in
ADD models \cite{Arkani-Hamed} for one extra dimension of
(effective) size $R_{ADD}=1/k$:
\begin{equation}
 M_{P}^2 = M_\star^{2+n} R_{ADD}^n.
 \label{rsmp}
\end{equation}

\subsection{The energy-momentum for RS models}

The Lagrangian density for Randall-Sundrum background can be
written as
\begin{equation}
{\cal{L}}={\cal{L}}_{g}+{\cal{L}}_{m},
\end{equation}
where ${\cal{L}}_{m}$ denotes the matter part and
\begin{eqnarray}
 {\cal{L}}_{g}&=&\sqrt{g} \left( \frac{1}{2k_\star^2} R
       +\Lambda\right) - \tau \sqrt{-g_{v}} \; \delta(y)
       + \tau \sqrt{-g_{h}} \; \delta(y-\pi r), \label{Lg1} \\
R~&=&(\omega_a \omega^a-\omega_{abc}\omega^{cba} )
    -\frac{2}{\sqrt{g}}\;{\partial}_{\mu}(\sqrt{g} \;
    e_{a}^{\mu}\omega^{a}), \\
 \omega_{abc}&=& \frac{1}{2} (\Omega_{abc}-\Omega_{bca}+\Omega_{cab}),\\
 \Omega_{abc}&=& e^{\mu}_{a} e^{\nu}_{b}({\partial}_{\mu} e_{c\nu}
               -{\partial}_{\nu} e_{c\mu}),\\
 \omega_{a}&=& \eta^{bc} \omega_{bac}=\omega^{c}_{\;\;ac}.
\end{eqnarray}
Eliminating the divergence expression, we rewrite the
${\cal{L}}_{g}$ as
\begin{eqnarray}\label{Lg2}
{\cal{L}}_{g}=\frac{1}{2k_\star^2} (\omega_a \omega^a
               -\omega_{abc} \omega^{cba} )\sqrt{g}
               + \Lambda \sqrt{g}
               - \tau\sqrt{-g_{v}} \; \delta(y)
               + \tau \sqrt{-g_{h}} \; \delta(y-{\pi} r).
\end{eqnarray}

For transformations Eq. (\ref{translation}), Eq.
(\ref{ConservationLaw2}) implies
\begin{equation} \label{ConservationLaw3}
{\partial}_{\mu}(\tilde{I}^{\mu}_{\sigma}e^{\sigma}_{a}+\tilde{V}^{\mu\nu}_{\sigma}
e^{\sigma}_{a,\nu})=0.
\end{equation}
From Einstein equations $\sqrt{g} \; T^{\mu}_{a}
=[{\cal{L}}_{g}]_{e^{a}_{\mu}}$ and Eq. (\ref{Imusigma}), we can
express $\tilde{I}^{\mu}_{\nu}e^{\nu}_{a}$ as
\begin{equation} \label{Imua}
\tilde{I}^{\mu}_{\nu}e^{\nu}_{a}=\left({\cal{L}}_{g}\delta^{\mu}_{\nu}-
\frac{{\partial}{\cal{L}}_{g}}{{\partial}e^{a}_{\lambda,
\mu}}e^{a}_{\lambda,\nu} \right) e^{\nu}_{a}+\sqrt{g} \;
T^{\mu}_{a}.
\end{equation}
Defining
\begin{equation} \label{tmua}
\sqrt{g} \; t^{\mu}_{a}=\left({\cal{L}}_{g}\delta^{\mu}_{\nu}-
\frac{{\partial}{\cal{L}}_{g}}{{\partial}e^{a}_{\lambda,
\mu}}e^{a}_{\lambda,\nu} \right) e^{\nu}_{a}+
\frac{{\partial}{\cal{L}}_{g}}{{\partial}e^{\nu}_{b,\mu}}
e^{\sigma}_{b}e^{\nu}_{a , \sigma},
\end{equation}
and considering Eq. (\ref{Vmusigmanu}), we then have
\begin{equation} \label{IeVe}
\tilde{I}^{\mu}_{\sigma} e^{\sigma}_{a} +
\tilde{V}^{\mu\nu}_{\sigma} e^{\sigma}_{a,\nu} =\sqrt{g} \;
(T^{\mu}_{a}+t^{\mu}_{a}).
\end{equation}
So Eq. (\ref{ConservationLaw3}) can be written as
\begin{equation} \label{ConservationLaw4}
{\partial}_{\mu}[\sqrt{g} \; (T^{\mu}_{a}+t^{\mu}_{a})]=0.
\end{equation}
This equation is the desired general covariant conservation law of
energy-momentum for Randall-Sundrum system. $t^{\mu}_{a}$ defined
in Eq. (\ref{tmua}) is the energy-momentum density of gravity
field, and $T^{\mu}_{a}$ to that of matter part. By virtue of Eq.
(\ref{Result2}), the expression on the LHS of Eq. (\ref{IeVe}) can
be expressed as divergence of superpotential $V^{\mu\nu}_{a}$
\begin{equation} \label{Tt}
\sqrt{g}
\;(T^{\mu}_{a}+t^{\mu}_{a})={\partial}_{\nu}V^{\mu\nu}_{a},
\end{equation}
where
\begin{equation} \label{SuperV}
V^{\mu\nu}_{a}=\tilde{V}^{\mu\nu}_{\sigma}e^{\sigma}_{a}=\frac{{\partial}{\cal{L}}_{g}}{{\partial}e^{\sigma}_{b,\mu}}e^{\nu}_{b}
e^{\sigma}_{a}.
\end{equation}
Eq. (\ref{Tt}) shows that the total energy-momentum density of a
gravity system always can be expressed as divergence of
superpotential. The total energy-momentum is
\begin{equation}\label{Pa}
P_{a}=\int_{\Sigma} d\Sigma_{\mu}
\sqrt{g}\;(T^{\mu}_{a}+t^{\mu}_{a})=\int_{S} dS_{\mu\nu}
V^{\mu\nu}_{a},
\end{equation}
where $dS_{\mu\nu}=(1/3!)\varepsilon_{\mu\nu\alpha\beta\gamma}
dx^{\alpha}\wedge dx^{\beta}\wedge dx^{\gamma}$.

Now we calculate the expressions of $V^{\mu\nu}_{a}$ and
$t^{\mu}_{a}$ by using the gravity Lagrangian density (\ref{Lg2})
of RS models. The explicit expressions are
\begin{eqnarray}
 V^{\mu\nu}_{a}&=&\frac{1}{k_{\star}^2}\left[e^{\mu}_{b} e^{\nu}_{c}
  \omega^{\;bc}_{a}
  + ( e^{\mu}_{a} e^{\nu}_{b} - e^{\nu}_{a} e^{\mu}_{b} )
  \omega^{b} \right] \sqrt{g},\label{VofRS} \\
t^{\mu}_{a}&=&  \frac{1}{2k_\star^2} \left\{
       e^{\mu}_{a}(\omega_b\omega^b-\omega_{cbd}\omega^{dbc})
       -2e^{\mu}_{b}(\omega_a\omega^b-\omega_{cad}\omega^{dbc}) \right. \nonumber\\
&&~~~~~~~\left.-2e^{\mu}_{c}(\omega_b\omega^{\;\;bc}_{a}+\omega_{b}\omega^{b\;\;c}_{\;\;a})
       + 2e^{\mu}_{d}\omega_{abc}\omega^{cbd} \right\} \nonumber\\
   &&+ e^{\mu}_{a} \left\{\Lambda-\tau\sqrt{\frac{-g_{v}}{g}} \;
       \delta(y) + \tau\sqrt{\frac{-g_{h}}{g}}\;\delta(y-\pi
       r)\right\}.
\end{eqnarray}
For the solution (\ref{rsmetric}), we can obtain the following
vierbein
\begin{equation} \label{vierbein}
e^{a}_{\mu} = (e^{-k \mid y
\mid}\delta^{\bar{a}}_{\mu},\delta^{4}_{\mu}). \;\;\;
(\bar{a}=0,1,2,3)
\end{equation}
The non-vanishing components of $\omega_{abc}$ and $\omega_{b}$
are
\begin{eqnarray} \label{omega_abc}
\omega_{004}=-k,  \;\; \omega_{114}= \omega_{224}= \omega_{334}= k
, \;\; \omega_{4}=4k.
\end{eqnarray}
For superpotential $V^{\mu\nu}_{a}$, the calculated result is
\begin{eqnarray} \label{V}
V^{04}_{0}= V^{14}_{1}= V^{24}_{2} = V^{34}_{3} =
-3\frac{k}{k_\star^2}e^{-3k \mid y \mid}.
\end{eqnarray}
Substituting Eq. (\ref{V}) into integral expression (\ref{Pa})
gives
\begin{eqnarray} \label{Pa_result1}
P_{a}=\left(3\frac{k}{k_\star^2}(1-e^{-3k\pi r}) v,0,0,0,0
\right),
\end{eqnarray}
where $v$ stands for the volume of usual three dimensional space
$M_3$
\begin{equation}
v=\int_{M_3} \! d^3 x.
\end{equation}
Substituting $k_\star^2 =\frac{1}{2}M_\star^{-3}$ and Eq.
(\ref{kMpM*}) into Eq. (\ref{Pa_result1}) yields
\begin{eqnarray} \label{Pa_result2}
P_{a}=\left(6\frac{M_\star^{6}}{M_P^{2}}(1-e^{-3k\pi
r})(1-e^{-2k\pi r}) v,0,0,0,0 \right).
\end{eqnarray}
So the momentum of RS system is vanishing, but the energy
\begin{eqnarray} \label{Pa_result2}
E=6\frac{M_\star^{6}}{M_P^{2}}(1-e^{-3k\pi r})(1-e^{-2k\pi r}) v
\end{eqnarray}
is infinite, which is caused by the gravity on the warped extra
dimension. When the radius $r$ of the extra dimension is taken the
limit $r \rightarrow 0$, i.e., the extra dimension disappears, the
bulk is just degenerates into our flat universe and the energy of
it vanishes. This result agrees with others.

From Eqs. (\ref{Tt}) and (\ref{V}), we can get the energy density
$\varepsilon$:
\begin{eqnarray} \label{EnergyDensity}
\varepsilon   = \varepsilon_0 e^{-3k \mid y
\mid}, \;\;\; %
\varepsilon_0 = \frac{9 M_\star^{9}}{2 M_P^{4}} (1-e^{-2k\pi r}).
\end{eqnarray}
Clearly, from this expression, it can be concluded that the energy
of the bulk distributes mainly near the Planck brane and force
lines are denser near this brane. This conclusion can be used to
explain why the gravitation on the SM brane (is our universe) is
very weak.

\section{Discussions}\label{RSConservationIV}

To summarize, by the use of general N\"{o}ether theorem, we have
obtained the conservation law of energy-momentum for the RS models
with the respect to the general displacement transform. The
energy-momentum current has a superpotential and are therefore
identically conserved. General covariance is a fundamental demand
for conservation laws in general relativity, and our definition
Eqs. (\ref{Tt}) and (\ref{Pa}) of energy-momentum is coordinate
independent. This conservation law is a covariant theory with
respect to the generalized coordinate transformations, but the
energy-momentum tensor is not covariant under the local Lorentz
transformation which, due to the equivalent principle, is
reasonable to require.


The conservative energy-momentum current and the corresponding
superpotential for the RS models are the same with those in (3+1)-
and (2+1)-dimensional Einstein theories, the Lagrangian density
${\cal{L}}_{h,v}$ corresponding to the hidden and visible brane
parts do not play a role in the conservation law, but they have an
influence to the energy of gravity. Both energy-momentum current
and the superpotential are determined only by vierbein field.

It is shown that, for the solution that gives a flat induced
metric on the branes, the momentum vanishes but the total energy
of bulk is infinite. This is quite different from the solution of
an isolated system in 3+1 dimension. The reason is that, though
the branes are flat, the extra dimension is warped even at the
infinite of them. When the extra dimension disappears, the bulk is
just degenerated into our four dimensional flat spacetime and the
energy vanishes. From the energy density formula
(\ref{EnergyDensity}), one can see that most of the bulk energy is
localized near the Planck brane, so the gravity on the SM brane is
very weak.

We think a conservation law of angular momentum
\cite{Liugr-qc/0508113} is also important in order to understand
the conservative quantities for the RS models. This conservation
law has been obtained using the approach in Ref.
\refcite{Duan1995}.

\section*{Acknowledgement}
It is a pleasure to thank Dr Liming Cao, Zhenhua Zhao and Zhenbin
Cao for interesting discussions. This work was supported by the
National Natural Science Foundation of the People's Republic of
China and the Fundamental Research Fund for Physics and Mathematic
of Lanzhou University.

\end{document}